# An Effective Descriptor for Predicting and Designing High-Temperature Ambient-Pressure Superconductors

Yuting Sun[1#], Shixu Liu[1#], Xin-Gao Gong[1]*, and Ji-Hui Yang[1]*

[1]Key Laboratory for Computational Physical Sciences (MOE), State Key Laboratory of Surface Physics, Department of Physics, Fudan University, Shanghai 200433, China

*corresponding authors
Email: xggong@fudan.edu.cn; jhyang04@fudan.edu.cn
#These authors contribute equally to this work.

## ABSTRACT:

Searching for ambient-pressure conventional superconductors with critical temperatures ($T_C$) higher than 40 K is a key challenge in the field of high-temperature superconductivity, mainly due to lack of efficient and effective models to estimate $T_C$ of potential systems. In this work, we propose a simplified model to estimate the dimensionless electron-phonon coupling (EPC) strength $\lambda$ by separately treating the EPC matrix elements which evaluate the pairing strength and the phonon-assisted nesting function $P(\omega)$ which evaluates the matching of electron bands and phonon spectra for forming potential electron pairs via phonons. Our model illuminates the critical role of $P(\omega)$ and its spectral integral $P$ in determining $\lambda$, i.e., high $P$ is a necessary condition leading to large $\lambda$ and thus high $T_C$, which is further demonstrated by showing that the reported high-$T_C$ traditional superconductors in literatures all have high $P$. As an easily quantifiable parameter, $P(\omega)$ and $P$ provide an efficient and effective descriptor for accelerating the discovery and rational design of high-$T_C$ superconductors. By applying the model to screen over the Computational 2D Materials Database (C2DB), we successfully identify several high-$T_C$ superconducting systems as confirmed by accurate first-principles calculations. Our model opens new avenues for exploring high-$T_C$ systems.

Superconductivity has been a highly prominent and intensely studied topic since its first discovery in mercury back to 1911 [1]. Despite more than a century of effort, superconductors with high $T_C$ at ambient pressure are still very rare in the field of condensed matter physics. So far, ambient-pressure superconductors with $T_C$ above 40 K are only limited to unconventional superconductors including copper oxides [2-4] and some iron-based compounds [5] while the record $T_C$ of conventional superconductors is only 39 K, which was reported in bulk $MgB_2$ in 2001 [6]. Note that they are all first discovered by experiments without exceptions. Theoretically, because mechanisms for unconventional superconductors are still unclear and thus reliable predictions are quite lacking, intensive efforts have been devoted to searching for conventional superconductors, which can be well explained by Bardeen-Cooper-Schrieffer (BCS) theory [7,8] and accurately predicted by Eliashberg methods [9,10]. Great success has been achieved by predicting high-$T_C$ hydride compounds such as $SH_3$ [11], $LaH_{10}$ [12], and $CaH_6$ [13] under extremely high pressures, which were later confirmed by experiments [14,15]. Meanwhile, several systems are predicted to have $T_C$ above 40 K at ambient pressure, such as $Al_4H$ (54 K [16]), $(Be_4)_2H$ (72-84 K [17]), hydrogenated monolayer $MgB_2$ (67 K [18]), $LiB_2C_2$ trilayer films (92 K [19]), $Li_3B_4C_2$ (54 K [20]) and $Mg_2XH_6$ (X= Rh, Ir, Pd, or Pt) (45-160 K [21,22]). These theoretical predictions represent significant advancements in our understanding of superconducting mechanisms and provide valuable guidance for experimental efforts. Continuing to explore and predict high-temperature ambient-pressure conventional superconductors remains a central priority in condensed matter physics, as such materials would combine fundamental scientific significance with transformative technological applications across energy, transportation, and computing sectors.

The prevailing method for predicting superconductors is solving Eliashberg equations, which is, however, very challenging due to heavy computational cost. Given the vast material space of potential superconducting compounds, with millions of possible chemical combinations and crystal structures, performing full calculations for all candidates becomes computationally prohibitive and economically unfeasible. This necessitates efficient preliminary screening methods to identify promising candidates before expensive ab initio calculations. It is generally accepted that systems with high density of states (DOS) at the Fermi surface $N(0)$ and high phonon frequency or Debye temperature $\theta_D$ tend to have high $T_C$ [23-25]. Actually, current high-throughput screenings of high-$T_C$ BCS superconductors using density functional theory (DFT) calculations and machine learning methods do often use these fundamental parameters in their models [23,24,26]. Some works also consider estimated EPC strength [27,28] in order to make accurate predictions. Besides, large Fermi surface nesting is also thought as possible indications of high $T_C$ [29]. Despite significant advances in the search for superconductors, development of an efficient model for preliminary screening of high-temperature ambient-pressure conventional superconductors remains an unmet challenge, limiting pace of discovery in this critical field.

In this work, we first derive a simplified model to estimate dimensionless EPC strength $\lambda$ to accelerate the search of high-$T_C$ superconductors. Within Migdal-Eliashberg theory framework, isotropic Eliashberg spectral function for phonon energy $\omega$ is given by [9,10,30]:

$$\alpha^2F(\omega) = \frac{1}{2N(0)} \sum_{mn\nu\mathbf{kq}} |g_{mn\nu}(\mathbf{k},\mathbf{q})|^2 \delta(\varepsilon_{n\mathbf{k}} - \varepsilon_{\mathrm{F}}) \delta(\varepsilon_{m\mathbf{k+q}} - \varepsilon_{\mathrm{F}}) \delta(\omega - \omega_{\mathbf{q}\nu}) \quad (1),$$

where $\varepsilon_{n\mathbf{k}}$ and $\varepsilon_{m\mathbf{k+q}}$ are eigenvalues for Kohn-Sham states, while $\varepsilon_{\mathrm{F}}$ refers to Fermi energy. $\omega_{\mathbf{q}\nu}$ represents frequency of phonon mode $\mathbf{q}\nu$. $g_{mn\nu}(\mathbf{k},\mathbf{q})$ is corresponding EPC matrix element. The dimensionless EPC strength $\lambda$ is then expressed as [31,32]:

$$\begin{aligned}\lambda &= 2\int_0^\infty \frac{\alpha^2F(\omega)}{\omega} \mathrm{d}\omega \\ &= \int_0^\infty \frac{1}{N(0)} \sum_{mn\nu\mathbf{kq}} \frac{1}{\omega} |g_{mn\nu}(\mathbf{k},\mathbf{q})|^2 \delta(\varepsilon_{n\mathbf{k}} - \varepsilon_{\mathrm{F}}) \delta(\varepsilon_{m\mathbf{k+q}} - \varepsilon_{\mathrm{F}}) \delta(\omega - \omega_{\mathbf{q}\nu}) \,\mathrm{d}\omega \quad (2).\end{aligned}$$

Once $\lambda$ is known, Allen-Dynes-modified McMillan formula gives [31,32]

$$T_{\mathrm{C}} = \frac{\omega_{\log}}{1.2} \exp\left[-\frac{1.04(1+\lambda)}{\lambda - \mu_c^*(1+0.62\lambda)}\right] \quad (3),$$

where $\mu_c^*$ is Coulomb pseudopotential with values ranging from 0.1 to 0.2. $\omega_{\log}$ is logarithmic average of phonon frequencies and can be calculated as:

$$\omega_{\log} = \exp\left[\frac{2}{\lambda}\int_0^\infty \frac{\alpha^2F(\omega)}{\omega} \log(\omega) \mathrm{d}\omega\right] \quad (4).$$

It is well known that obtaining accurate converged $\lambda$ and thus superconductivity requires EPC matrix elements on very dense $\mathbf{k}$ and $\mathbf{q}$ grids, demanding very expensive computational costs that hinder efficient prediction of high-$T_C$ superconductors. To avoid heavy EPC calculations, here we consider phonon-energy resolved EPC strength $\frac{\partial\lambda}{\partial\omega}$, which satisfies $\lambda = \int_0^\infty \frac{\partial\lambda}{\partial\omega} \mathrm{d}\omega$ and can be written as:

$$\frac{\partial\lambda}{\partial\omega}=\frac{1}{N(0)}\sum_{mn\nu\mathbf{kq}}\frac{1}{\omega}|g_{mn\nu}(\mathbf{k},\mathbf{q})|^2\delta(\varepsilon_{n\mathbf{k}}-\varepsilon_{\mathrm{F}})\delta(\varepsilon_{m\mathbf{k+q}}-\varepsilon_{\mathrm{F}})\delta(\omega-\omega_{\mathbf{q}\nu}) \quad (5).$$

Further, we decompose $\frac{\partial\lambda}{\partial\omega}$ into two separate terms, i.e.:

$$\frac{\partial\lambda}{\partial\omega}=|g^*(\omega)|^2P(\omega) \quad (6),$$

where $g^*(\omega)$ can be seen as phonon-energy resolved and wavevector-averaged EPC matrix element. $P(\omega)$ is expressed as:

$$P(\omega)=\frac{1}{N(0)}\sum_{mn\nu\mathbf{kq}}\frac{1}{\omega}\delta(\varepsilon_{n\mathbf{k}}-\varepsilon_{\mathrm{F}})\delta(\varepsilon_{m\mathbf{k+q}}-\varepsilon_{\mathrm{F}})\delta(\omega-\omega_{\mathbf{q}\nu}) \quad (7),$$

which characterizes the potential for two electrons with energies $\varepsilon_{n\mathbf{k}}$ and $\varepsilon_{m\mathbf{k+q}}$ to form a Cooper pair through exchange of a phonon with energy $\omega$. Compared to Fermi surface nesting function, $P(\omega)$ has an extra $\delta(\omega-\omega_{\mathbf{q}\nu})$ term which specifically includes roles of phonons. Therefore, we name $P(\omega)$ as phonon-assisted nesting function which can be easily and accurately calculated from band structure and phonon spectrum. Note that, a larger integration of $P(\omega)$ over $\omega$, i.e., $P=\int P(\omega)\,\mathrm{d}\omega$, indicates that electrons have larger chance to form pairs via phonons and therefore a potentially higher $T_C$. Considering that, larger systems, such as supercells, typically contain more electrons per cell, it is more appropriate to normalize $P$, i.e., by dividing it by total atomic mass $M$ of the cell.

As for the $g^*(\omega)$ term, although it is difficult to get accurate results without performing first-principles calculations, some knowledge learned from previous studies can still be borrowed. For example, it is known that there is isotope effect in BCS superconductors which suggests if the element involved in forming Cooper pairs has larger atomic mass $m^*$ (typically the element has significant contributions to the electronic density of states at the Fermi level), it often leads to smaller EPC, i.e.:

$$|g^*(\omega)|^2\propto\frac{1}{m^*} \quad (8).$$

Combining Eqs. (5) and (6), we can generally use $\frac{P}{m^*M}$ as a preliminary screening descriptor for large $\lambda$ and thus high $T_C$. Details can be found in the Supplemental Materials (SM).

To systematically validate the effectiveness of using $\frac{P}{m^*M}$ as a reliable descriptor, we calculate electronic band structures and phonon spectra of known superconductors from both experimental observations and theoretical predictions in literature [6,16,18,21,23,33-48], covering a variety of systems including monolayers and bulks with diverse structures and symmetries. Without performing heavy EPC calculations, $\frac{P}{m^*M}$ and $P(\omega)$ can be easily obtained from Eq. 7. Figure 1 shows the correlation between our calculated $\frac{P}{m^*M}$ values and literature-reported $T_C$ values. Note that, although it is difficult to build a quantitative relationship between $\frac{P}{m^*M}$ and $T_C$ as the $|g^*(\omega)|^2$ term is very sensitive across different systems, but qualitatively, larger $\frac{P}{m^*M}$ is beneficial for high $T_C$. As seen, using $\frac{P}{m^*M}$ value of bulk $MgB_2$ as a benchmark criterion, superconductors exhibiting higher $T_C$ values than $MgB_2$'s, such as $Mg_2IrH_6$, metal borocarbides, and hydrogenated monolayers, consistently demonstrate larger $\frac{P}{m^*M}$ than $MgB_2$'s. Conversely, systems characterized by significantly small $\frac{P}{m^*M}$ such as two-dimensional (2D) $BeP_2$ [33], 2D $Ba_2N$ [34], 2D $Ca_2N$ [35], 2D $Zr_2S_2$ [23], bulk LaO [36,37], and bulk ZrN [38] cannot achieve comparable high-temperature superconductivity. For 2D $MgB_2$, although it has a higher $\frac{P}{m^*M}$ value, its $T_C$ is lower than $MgB_2$'s, but 20 K is still a relatively high $T_C$ among known ambient-pressure superconductors. Nevertheless, if one uses $\frac{P}{m^*M}$ as a predictive descriptor, one can easily identify all the high-$T_C$ systems by comparing $\frac{P}{m^*M}$ values with that of $MgB_2$, justifying that large $\frac{P}{m^*M}$ is a necessary condition for high-$T_C$ conventional superconductors.

Having established that our descriptor effectively identifies high-$T_C$ superconductors, a critical question naturally arises: what are the fundamental microscopic origins that give rise to enhanced $P$ and $P(\omega)$ values in these promising materials? To address this question systematically and provide mechanistic insights beyond empirical correlations, we perform in-depth comparative analyses. Specifically, we examine $\frac{\partial\lambda}{\partial\omega}$, $P(\omega)$, and $|g^*(\omega)|^2$ distributions, alongside isotropic superconducting properties, for two strategically selected high-$T_C$ systems: hydrogenated 2D $MgB_2$ ($HMgB_2$) [18] and three-dimensional (3D) $Mg_2IrH_6$ [21], benchmarked against the well-characterized $MgB_2$ [6]. The choice of these systems is deliberate—they represent distinct structural prototypes (2D vs. 3D) while both achieving superconductivity above 20 K, thus offering complementary perspectives on the universal and material-specific factors governing high $P$ and $P(\omega)$. Through this comparative analysis, we aim to elucidate the interplay between electronic structure, phonon dynamics, and EPC that ultimately determines the magnitude of our descriptor and, consequently, the superconducting transition temperature.

As seen in Fig. 2a, both $HMgB_2$ and $Mg_2IrH_6$ exhibit significantly larger $\lambda$ values than $MgB_2$. From detailed phonon-energy resolved spectra, we can find that, $P(\omega)$ of $HMgB_2$ and $Mg_2IrH_6$ as well as $P$ exhibit values that are several-fold or even several orders of magnitude higher than those of $MgB_2$, indicating that the matching of electron bands and phonon spectra in $HMgB_2$ and $Mg_2IrH_6$ is more favorable for the potential formation of Cooper pairs (see Fig. 2b). This can be attributed to two reasons. First, $N(0)$ is significantly higher in $HMgB_2$ and $Mg_2IrH_6$ and thus there are more electrons ready to pair. For example, $HMgB_2$ exhibits an $N(0)$ approximately 50% higher than $MgB_2$, while $Mg_2IrH_6$ shows nearly 4-fold greater values, indicating more electrons available at the Fermi level for pairing (see Fig. S1). Second, both $HMgB_2$ and $Mg_2IrH_6$ exhibit an abundance of phonons in relatively low-frequency region (see Fig. S2), though for different reasons. In $HMgB_2$, the hydrogenation process slightly stretches the Mg-B bond lengths, resulting in phonon softening. In $Mg_2IrH_6$, the presence of heavy Ir atoms naturally contributes to a greater number of low-energy phonon modes. In addition, H also contributes significant phonon density of states around 50~70 meV. The large $N(0)$ and large low-energy phonon density of states in $HMgB_2$ and $Mg_2IrH_6$ favor the matching between electronic band structures and the phonon spectra, especially in phonon energy region below 70 meV, leading to much larger $P(\omega)$ and thus $P$, which facilitates superconducting pairing. It is worth mentioning that, having a large $N(0)$ or a large soft phonon density of states, or both, do not necessarily guarantee large $P(\omega)$ and $P$ values. For example, while the DOS of $CuH_2$ is suppressed to approximately half that of $MgB_2$, the peak maximum of $P(\omega)$ and $\frac{P}{m^*M}$ for $CuH_2$ exceeds those of $MgB_2$ by several times (see Fig. 1 and Fig. S19). Only through calculations using our defined Eq. 7 can one accurately determine whether a system possesses large $P(\omega)$ and $P$.

While the dramatically large $P(\omega)$ explains part of superior $\lambda$ in $HMgB_2$ and $Mg_2IrH_6$, our decomposition for $\lambda$ in Eq. 6 indicates that the pairing strength $|g^*(\omega)|^2$ also play a role. Note that, the paring strength $|g^*(\omega)|^2$ in both $HMgB_2$ and $Mg_2IrH_6$ keeps same order of magnitude as that in $MgB_2$ in phonon energy region below 50 meV (see Fig. 2c). Together with the large $P(\omega)$, the integrated $\frac{\partial\lambda}{\partial\omega}$ in $HMgB_2$ and $Mg_2IrH_6$ is much larger than that in $MgB_2$ (see Fig. 2a) in low-energy phonon region. In $MgB_2$, $\frac{\partial\lambda}{\partial\omega}$ only shows significant increase around the phonon energy region of about 60 meV because $|g^*(\omega)|^2$ is about one order of magnitude higher. Nevertheless, tremendously large $P(\omega)$ in $HMgB_2$ and $Mg_2IrH_6$ dominates over large paring strength in $MgB_2$, leading to overall larger $\lambda$ values. This observation further justifies the use of $P(\omega)$ for searching for high-$T_C$ superconductors.

Another interesting finding is that peaks of $P(\omega)$ and $|g^*(\omega)|^2$ are better matched in $HMgB_2$ and $Mg_2IrH_6$ while they are mismatched in $MgB_2$. If $MgB_2$ can be engineered to have better matching between its electronic band structure and phonon spectrum, a greatly enhanced $T_C$ could be expected. This observation provides insights into strategies for superconductivity optimization. If a superconductor can be engineered to enhance intrinsic $P(\omega)$, or to achieve optimal alignment between the spectral peaks of $P(\omega)$ and $|g^*(\omega)|^2$, a significantly enhanced $T_C$ would be expected. Taking $HMgB_2$ as an illustrative case, its $P$ value is several times greater than that of $MgB_2$. Meanwhile, $HMgB_2$ exhibits $|g^*(\omega)|^2$ peaks in low-$\omega$ region that correlate better with corresponding $P(\omega)$ distribution. These combined characteristics demonstrate that strategic modulation of $P(\omega)$ represents a promising approach for enhancing superconducting properties. It is worth noting that strengthening $|g^*(\omega)|^2$ is equally crucial for improving superconducting performance. As proven in previous studies [49], one important approach to modulate $|g^*(\omega)|^2$ is boosting the coupling of $\sigma$-bonding electrons with interlayer covalent bond-stretching modes, which presents an alternative pathway for achieving high-$T_C$ superconductivity. Together, these strategies can be synergistically employed to achieve superior superconducting performance.

To further demonstrate the practical application of $P$ and $P(\omega)$, we apply the above model to search for high-$T_C$ ambient-pressure superconductors. Here we choose the C2DB [50,51] for high-throughput screening because it contains information of phonon frequencies for many metallic systems. The workflow is shown in Fig. 3a. The starting point is the 16789 monolayer materials currently available in C2DB. First, we select the materials that meet the criteria of metallic behavior, dynamic stability, and non-magnetism, resulting in 1036 candidates. Next, considering the fact that low frequency will limit $\omega_{\log}$ and thus $T_C$, we require maximum phonon frequency $\omega_{\max} \geq 42$ meV considering that exponential term in Eq. 3 is usually no larger than 0.1~0.15 at ambient pressure, resulting in 576 candidates. Then, because phonon spectra are not available for many systems and accurately computing them is computationally demanding, we performed first-principles phonon calculations using Vienna ab initio simulation package (VASP) [52,53] for these 576 candidates to assess their dynamical stability. Considering the computational cost of phonon calculations in large-scale high-throughput workflows, these evaluations were conducted with moderately converged settings. The complete phonon spectra of the dynamically stable candidates are provided in SM where the corresponding C2DB IDs are also indicated. Note that, this step can be significantly accelerated if band structures and phonon spectra can be easily obtained, i.e., using machine-learning methods [54,55]. Subsequently, we conduct first-principles calculations of band structures and phonon spectra to obtain accurate $P(\omega)$ and $P$. The systems with imaginary phonon modes are spontaneously discarded. As discussed above, larger $\frac{P}{m^*M}$ indicates larger $\lambda$. To further narrow the number of ideal candidates, we also consider that, if $P$ has a larger portion in high-$\omega$ region, more optical phonon modes will be involved in forming Cooper pairs and thus leading to larger $\omega_{\log}$. Therefore, we would expect systems lying in the top right corner of Fig. 3b tend to have high $T_C$. Finally, we use EPW code [56-58] to accurately calculate their superconductivity properties (see SM).

After performing the above high-throughput screening process, we successfully identify 6 materials with $T_C$>20 K according to accurate first-principles calculations. Their structures and $T_C$ values are shown in the inset of Fig. 3b. Among them, we find that $Ti_3N_2H_2$ has a $T_C$ value of 52 K (78 K) by solving isotropic (anisotropic) Migdal-Eliashberg equations on an interpolated k-point grid of 600×600×1 (300×300×1) and a q-point grid of 200×200×1 (100×100×1). Thorough convergence tests are carefully conducted with details provided in SM. In the following, we will just use $Ti_3N_2H_2$ as an example to illustrate the remarkable efficiency and predictive power of our approach in accelerating the identification of high-$T_C$ superconductors. Experimental verification of its superconductivity is out of the scope of this work. Fig. 4 presents its crystal structure, band structure with corresponding total DOS (TDOS) and orbital-projected DOS (PDOS), vibrational spectrum with corresponding phonon DOS (phDOS) and isotropic Eliashberg spectral function $\alpha^2F(\omega)$, as well as superconducting properties. As shown in Fig. 4(b), $N(0)$ of $Ti_3N_2H_2$ has a relatively large value of 11.2 states/eV that indicates more electrons available for pairing, which, together with more soft phonons from heavier Ti and N atoms, results in $P(\omega)$ and $P$ values 20-30 times higher than those of $MgB_2$. Further discussion of its outstanding superconducting properties can be found in SM.

The remarkable superconducting properties of $Ti_3N_2H_2$ become even more significant when placed in the broader context of MXene materials, among which $Ti_3N_2H_2$ has the highest $T_C$ reported to date as far as we know [59-61]. Moreover, MXenes have well-established preparation methods with more than 50 systems synthesized experimentally [62]. Specifically, $Ti_4N_3$ is successfully fabricated in 2016 [63], exfoliating the top TiN layer of which yields $Ti_3N_2$. Our calculated exfoliation energy of 73 meV/Å$^2$ is lower than successfully synthesized 2D materials like SnSe [64,65] (89 meV/Å$^2$ [66]) and hematene [67] (140 meV/Å$^2$ [68,69]), indicating its feasible synthesis. Then $Ti_3N_2H_2$ can be obtained by hydrogenation—an exothermic process releasing 2.53 eV per formula. Molecular dynamics (MD) simulations confirm H-Ti bonding stability at T=600K, indicating its robust thermodynamic stability (see SM). Thus, experimental realization of $Ti_3N_2H_2$ might be expected. Although currently confined to theoretical exploration, the rapid identification of this promising superconducting material through our framework demonstrates remarkable efficiency and predictive power of our approach. This success underscores the potentiality of our model for accelerating the discovery of novel high-temperature superconductors, even as experimental confirmation remains the next crucial step.

We also note that, $Ti_3N_2H_2$ has a more pronounced peak in DOS lying at about 0.2 eV below the Fermi level. Based on the $P(\omega)$ modulation approach discussed above, downward shifting of the Fermi level via hole doping would yield larger $N(0)$ and thus larger $P(\omega)$. This can be realized by replacing middle Ti layer with Sc layer. As shown in Fig. 4(c), middle metal layer has negligible contribution to states near the Fermi level, so newly designed $Ti_2ScN_2H_2$ system has similar band characters with $Ti_3N_2H_2$ except that it has more electrons ready to pair. Consequently, $Ti_2ScN_2H_2$ has even larger $P(\omega)$ and thus higher $T_C$. Our calculations show that $Ti_2ScN_2H_2$ remains dynamically stable and demonstrates remarkable superconductivity, with $T_C$ reaching 54 K (80 K) by employing isotropic (anisotropic) Migdal-Eliashberg theoretical frameworks (see SM).

Finally, using our proposed model, we can also understand why high-$T_C$ BCS superconductors are often theoretically reported in 2D or layered systems [6,49]. Generally speaking, on the one hand, 2D metals will have larger electron and phonon density of states due to quantum confinement effects [70]. Therefore, electrons have a larger chance to form pairs via phonons, that is, $P(\omega)$ tends to be larger in 2D. On the other hand, due to the reduced dimension, the screening effects are suppressed. Consequently, the pairing strength $|g^*(\omega)|^2$ will be increased. These two characters of 2D metals will help obtain larger $\lambda$, resulting in high $T_C$. Extending this argument, if a 3D superconductor can achieve sufficiently large $P(\omega)$ that rivals or even exceeds that of most 2D systems, it is highly likely to exhibit remarkable superconductivity, such as $Mg_2IrH_6$ mentioned above.

In summary, we have derived a simplified model to estimate $\lambda$ by separately treating EPC matrix elements which evaluate the pairing strength and the phonon-assisted nesting function $P(\omega)$ which evaluates potentiality of electron pairing via phonons, thereby greatly accelerating the search for high-$T_C$ superconductors. Based on this theoretical framework, we comprehensively elucidate the origin of the exceptional superconducting properties of ambient-pressure high-temperature superconductors reported in recent literature, particularly $HMgB_2$ and $Mg_2IrH_6$. By applying the model in a high-throughput workflow, we have identified high-$T_C$ ambient-pressure superconducting systems. Besides, we have explained why high-$T_C$ superconductors are often predicted in 2D systems using our model. We expect our work will help find more high-$T_C$ ambient-pressure superconductors in the future.

## ACKNOWLEDGMENTS

This work was partially supported by the China National Key R&D Program (2022YFA1404603), the National Natural Science Foundation of China (Grants No. 12188101, and No. 12474222), Innovation Program for Quantum Science and Technology (2024ZD0300102), Shanghai Science and Technology Program (Grant No. 23JC1400901), and the Guangdong Major Project of the Basic and Applied Basic Research (Future functional materials under extreme conditions 2021B0301030005).

## FIGURES

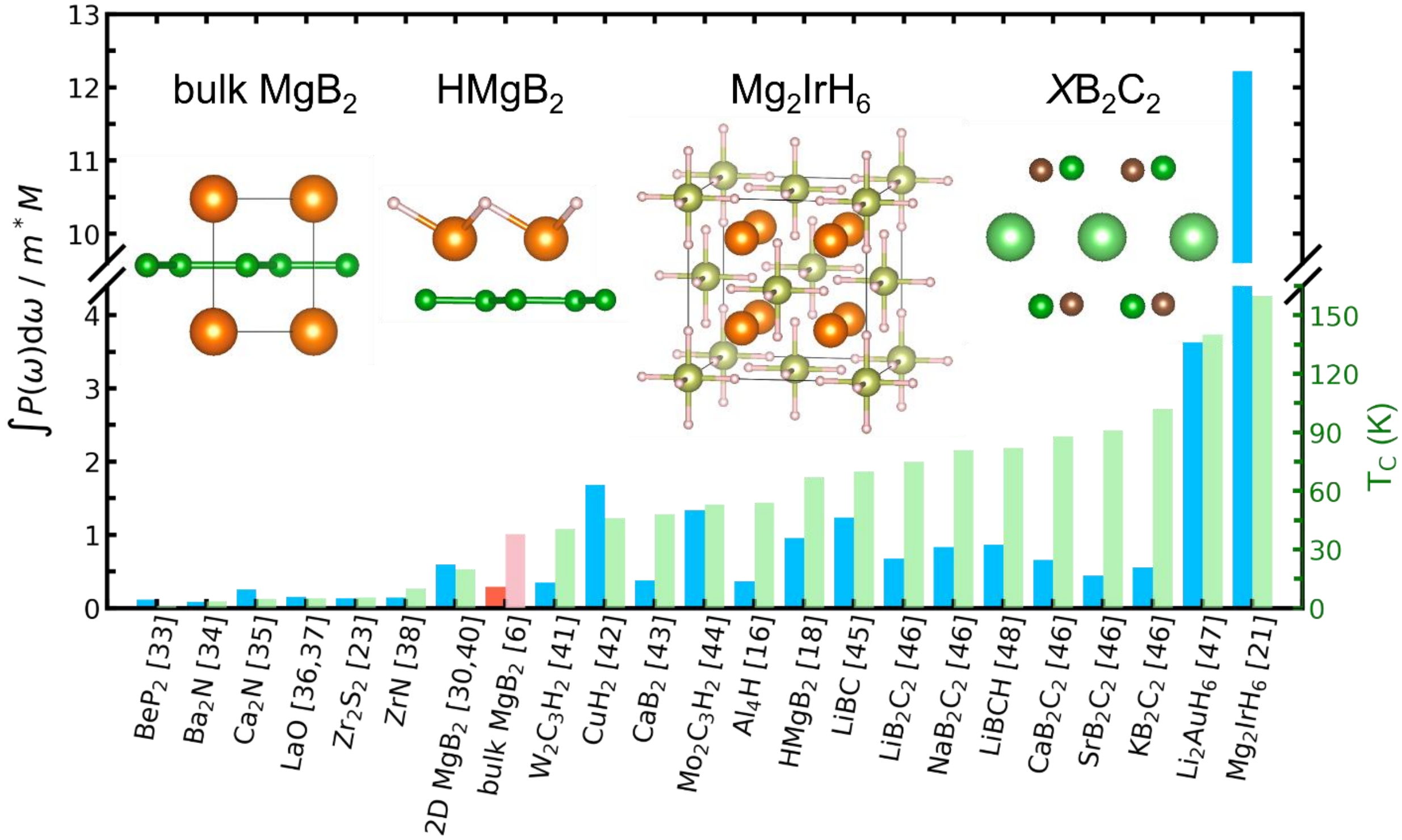

**Fig. 1.** The correlation between our calculated $\frac{P}{m^*M}$ values and literature-reported $T_C$ values. The structures of representative superconductors are shown in the inset. Here, the blue bars represent the $\frac{P}{m^*M}$ values of the materials, while the green bars represent their $T_C$ values. For $MgB_2$, which serves as a benchmark, its $\frac{P}{m^*M}$ value and $T_C$ are marked in red and pink, respectively. This validates $\frac{P}{m^*M}$ as an effective preliminary screening descriptor for identifying high-$T_C$ conventional superconductors.

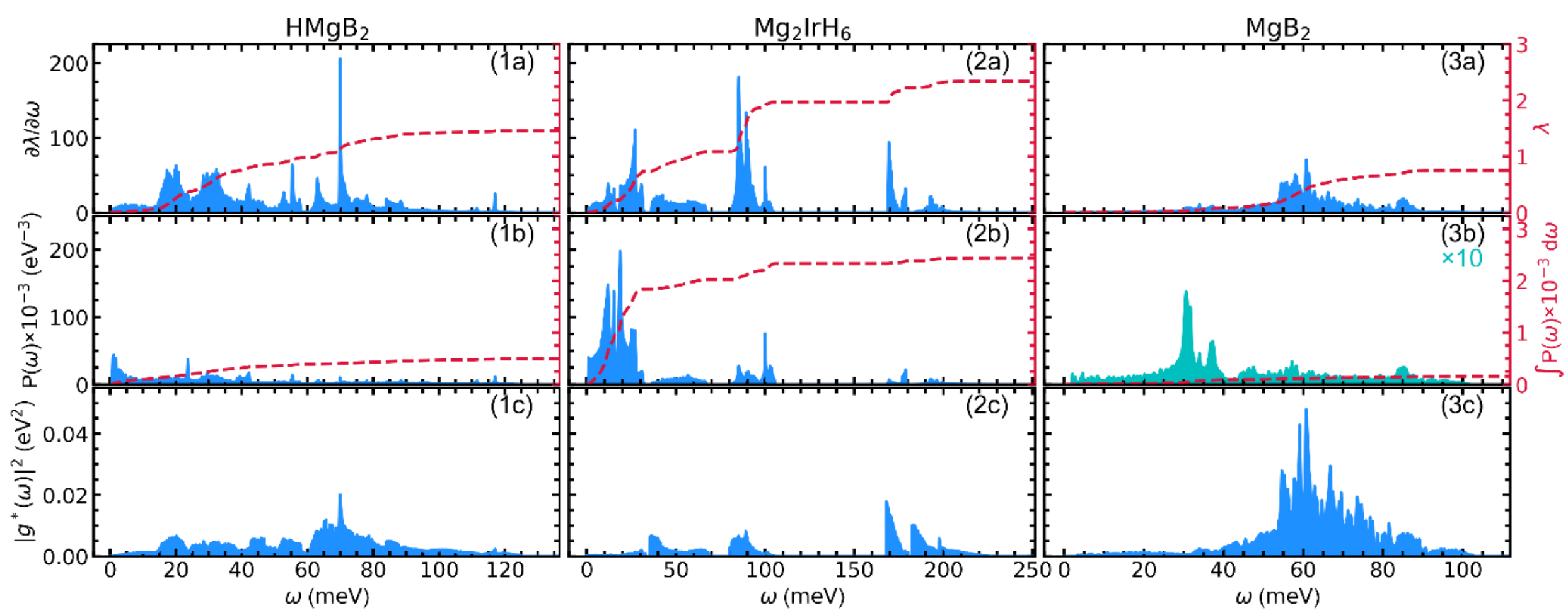

**Fig. 2.** Comparisons of superconductivity related properties for 2D $HMgB_2$, 3D $Mg_2IrH_6$ and bulk $MgB_2$. (a) The phonon-energy resolved dimensionless EPC strength $\frac{\partial\lambda}{\partial\omega}$, (b) phonon-assisted nesting function $P(\omega)$, and (c) averaged electron-phonon coupling $|g^*(\omega)|^2$ in $HMgB_2$, $Mg_2IrH_6$ and $MgB_2$. The red lines in the top panels indicate the integrated $\frac{\partial\lambda}{\partial\omega}$ over $\omega$, while those in the second panels indicate the integrated $P(\omega)$ over $\omega$. For better visibility, the corresponding $P(\omega)$ of $MgB_2$ has been amplified by a factor of 10 and marked in green. Note that, $P(\omega)$ of $HMgB_2$ and $Mg_2IrH_6$ as well as $P$ exhibit values that are several-fold or even several orders of magnitude higher than those of $MgB_2$, indicating that the matching of electron bands and phonon spectra in $HMgB_2$ and Mg2IrH$_6$ is more favorable for the potential formation of Cooper pairs, therefore resulting in their significantly larger $\lambda$ values.

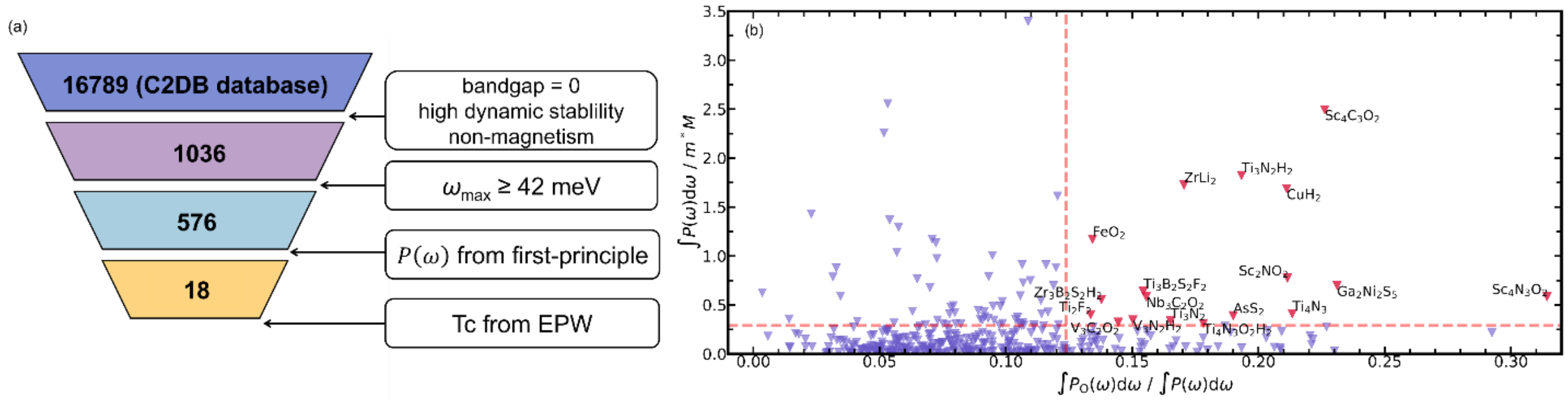

**Fig. 3.** Schematics illustrating the screening process of identifying high-temperature ambient-pressure 2D superconductors. (a) An illustration of preliminary high-throughput workflow applied to the C2DB database. The screening criteria and number of materials at each step of the workflow are given. (b) Fine screening of candidate materials based on $\frac{P}{m^*M}$ and portion of $P$ in high-$\omega$ region denoted as $P_O$, both obtained from first-principles calculations of electron structures and phonon spectra. Two red lines oriented perpendicular and parallel to the x-axis represent threshold values obtained from calculations for $MgB_2$. Our model allows screening thousands of candidates efficiently and successfully identifies multiple materials with $T_C$ higher than 20K.

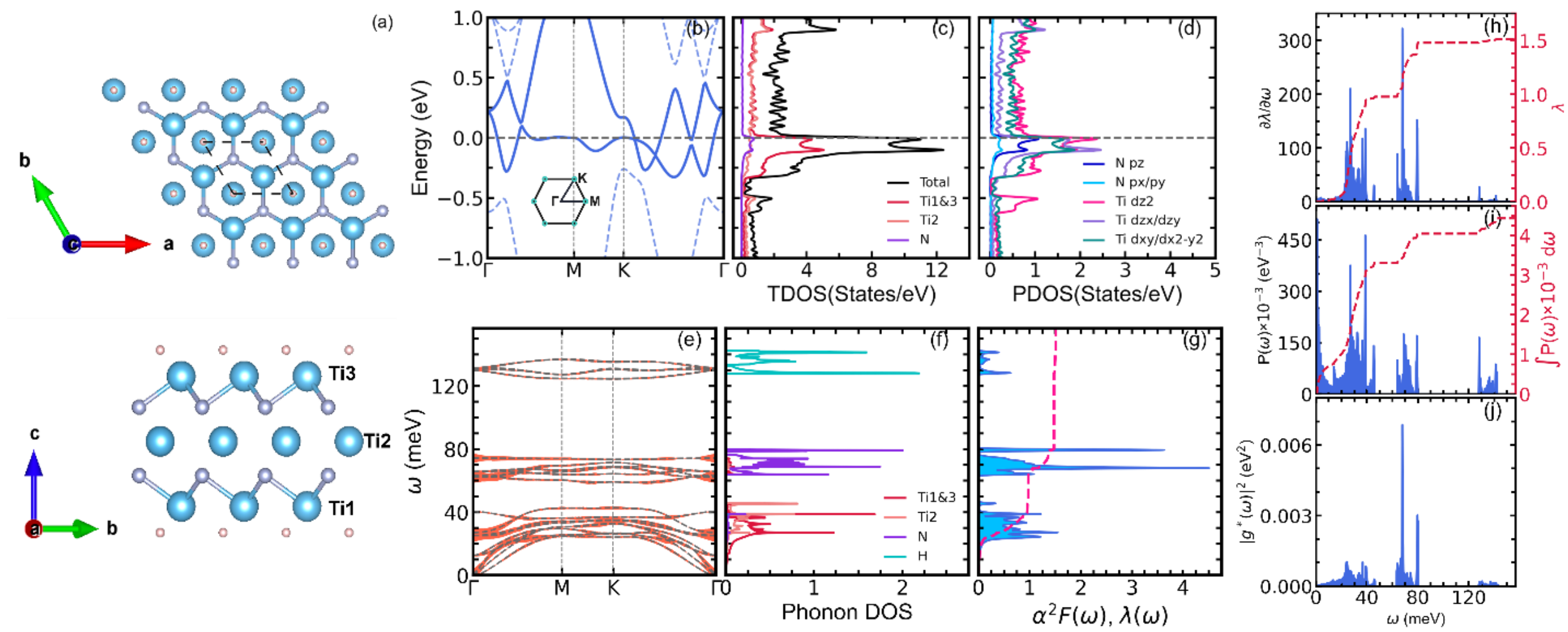

**Fig. 4.** Structures, electronic structures, vibrational and superconducting properties of $Ti_3N_2H_2$ using LDA functionals. (a) Top and side views of $Ti_3N_2H_2$ monolayer. Titanium, nitrogen and hydrogen atoms are represented by blue, purple and pink spheres, respectively. (b) Electronic band structure of $Ti_3N_2H_2$ along high-symmetry line Γ-M-K-Γ. The Fermi level indicated by the dotted line is set to 0 eV. (c) The total DOS (TDOS) of $Ti_3N_2H_2$ contributions of Ti, N and H atoms. (d) The partial density of states (PDOS) of $Ti_3N_2H_2$. (e) Phonon dispersion of $Ti_3N_2H_2$ weighted by the magnitude of the EPC constant $\lambda_{\mathbf{q}\nu}$ for phonon mode $\mathbf{q}\nu$. (f) The atomic projected phonon DOS. (g) Eliashberg spectral function $\alpha^2F(\omega)$ and cumulative electron-phonon coupling parameter $\lambda(\omega)$. (h) The phonon-energy resolved dimensionless EPC strength $\frac{\partial\lambda}{\partial\omega}$, (i) phonon-assisted nesting function $P(\omega)$, and (j) averaged electron-phonon coupling $|g^*(\omega)|^2$. The orange lines in the top panels indicate the integrated $\frac{\partial\lambda}{\partial\omega}$ over $\omega$, which is also equal to $\lambda(\omega)$ in Fig. 4g. The orange lines in the second panels indicate the integrated $P(\omega)$ over $\omega$.